\documentclass{emulateapj}
\usepackage{amsmath}
\usepackage{txfonts}
\usepackage{natbib}
\usepackage{graphicx}
\usepackage{color}
\usepackage{multirow}
\usepackage{array}

\newcommand{\ud}{\ensuremath{\mathrm{d}}}

\shorttitle{Neutrino-driven supernova of a low-mass progenitor in three dimensions}

\shortauthors{Melson, Janka, \& Marek}

\begin{document}

\title{Neutrino-driven supernova of a low-mass iron-core progenitor\\ 
boosted by three-dimensional turbulent convection}

\author{Tobias Melson\altaffilmark{1,2}, Hans-Thomas Janka\altaffilmark{1}, 
and Andreas Marek\altaffilmark{3}}

\altaffiltext{1}{Max-Planck-Institut f\"ur Astrophysik, Karl-Schwarzschild-Str.~1, 
85748 Garching, Germany}
\altaffiltext{2}{Physik Department, Technische Universit\"at M\"unchen, 
James-Franck-Stra\ss e 1, 85748 Garching, Germany}
\altaffiltext{3}{Rechenzentrum der Max-Planck-Gesellschaft (RZG), Boltzmannstr. 2, 
85748 Garching, Germany}

\begin{abstract}
We present the first successful simulation of a neutrino-driven supernova explosion
in three dimensions (3D), using the \textsc{Prometheus-Vertex} code with an axis-free
Yin-Yang grid and a sophisticated treatment of three-flavor, energy-dependent
neutrino transport. The progenitor is a nonrotating, zero-metallicity 9.6\,$M_\odot$
star with an iron core. While in spherical symmetry outward shock acceleration sets 
in later than 300\,ms after bounce, a successful explosion starts at $\sim$130\,ms
postbounce in two dimensions (2D). The 3D model explodes at about the same time but
with faster shock expansion than in 2D and a more quickly increasing and roughly
10\% higher explosion energy of $>$10$^{50}$\,erg. The more favorable explosion 
conditions in 3D are explained by lower temperatures and thus reduced neutrino
emission in the cooling layer below the gain radius. This moves the gain
radius inward and leads to a bigger mass in the gain layer, whose larger recombination 
energy boosts the explosion energy in 3D. These differences are caused by less coherent, 
less massive, and less rapid convective downdrafts associated with postshock convection
in 3D. The less violent impact of these accretion downflows in the cooling layer
produces less shock heating and therefore diminishes energy losses by neutrino
emission. We thus have, for the first time, identified a reduced mass accretion
rate, lower infall velocities, and a smaller surface filling factor of convective 
downdrafts as consequences of 3D postshock turbulence that facilitate 
neutrino-driven explosions and strengthen them compared to the 2D case.
\end{abstract}

\keywords{
supernovae: general --- hydrodynamics --- instabilities --- neutrinos}

\section{Introduction}

Modeling the core-collapse supernova (SN) mechanism in three dimensions (3D) 
is still in its infancy. Generalization from spherical symmetry (1D) to
axial symmetry (2D) introduces nonradial flows and hydrodynamic instabilities
like convection and the standing accretion shock instability (``SASI''; 
\citealp{blondin_03}) in the neutrino-heated postshock layer, which have been
recognized as helpful for the explosion due to improved neutrino-heating
conditions and buoyancy or turbulent pressure behind the shock (e.g., 
\citealp{herant_94,burrows_95,janka_96,murphy_08,murphy_13,marek_09,mueller_12a,mueller_12b,mueller_13,mueller_14a,mueller_14b,suwa_13,bruenn_13,bruenn_14,takiwaki_14,couch_14o}).

Explosions of 2D models, however, start late and their energies 
tend to be fairly low
(except for those of \citealp{bruenn_13,bruenn_14}, whose results still need 
better explanation and confirmation by detailed comparisons with other recent
simulations). 3D effects were hoped to improve the situation, but a first
optimistic report based on parametrically triggered neutrino-driven explosions
by \cite{nordhaus_10} was not supported by subsequent works
\citep{hanke_12,couch_13,couch_14,takiwaki_12,takiwaki_14,mezzacappa_15}.
\cite{dolence_13} and \cite{burrows_12}
confessed an error in \cite{nordhaus_10}, nevertheless they still claimed a
remaining positive effect in 3D. Turbulent fragmentation by energy cascading from
large to small scales, however, seems to disfavor and delay explosions in 3D,
because it destroys the biggest convective plumes, which were recognized
as helful for 2D explosions \citep{hanke_12,couch_13,couch_14,dolence_13}. 
Therefore \cite{abdikamalov_14} are worried about even less
favorable explosion conditions when resolution shortcomings of all current 
3D models can be overcome, and \cite{couch_13o,couch_14o} advocate 
pre-collapse progenitor-core asymmetries as possible solution of the dilemma.

Here we present the first successful 3D simulation of a neutrino-driven SN 
explosion of a 9.6\,$M_\odot$
iron-core star computed fully self-consistently with the neutrino-hydrodynamics
code \textsc{Prometheus-Vertex}. We identify, for the first time, consequences
of 3D turbulence in the convective gain layer that enhance the explosion 
energy and accelerate shock expansion in 3D relative to 2D.
While so far in more massive progenitors 3D turbulence appears less supportive
for the initiation of explosions than 2D flows (cf.\ references above and
\citealp{hanke_13,tamborra_14} for 11.2,\,20,\,27\,$M_\odot$ models), 
we investigate here whether 3D can strengthen the explosion {\em after}
the onset of shock runaway. The considered 9.6\,$M_\odot$ model offers an
ideal case for this study because it explodes at the same time in 2D and 3D.
We briefly describe our numerical approach in Sect.~\ref{sec:numerics},
discuss our results in Sect.~\ref{sec:results}, and conclude in  
Sect.~\ref{sec:conclusions}.

\begin{figure*}[t!]
\begin{center}
\includegraphics[width=.33\textwidth]{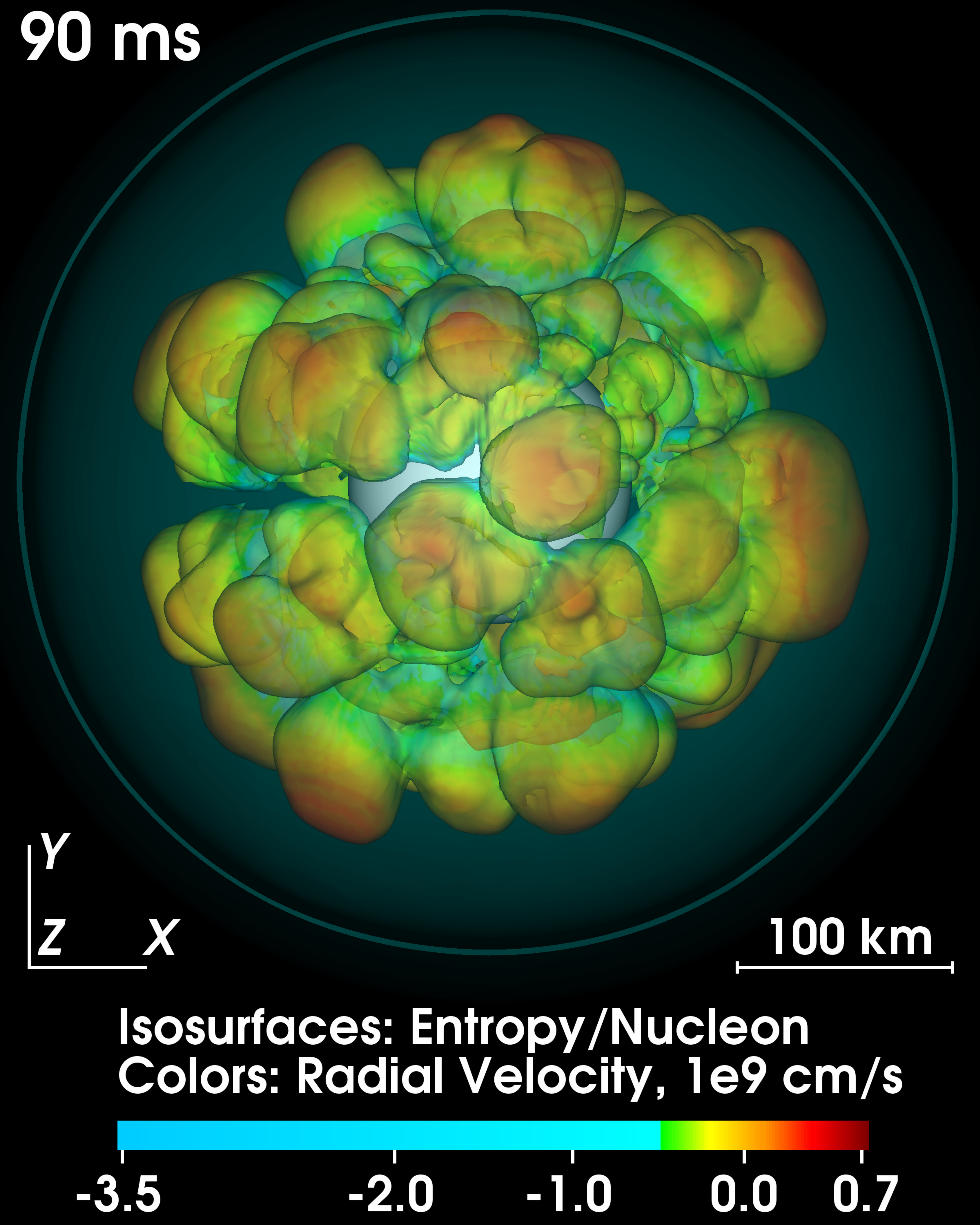}\hskip1.0pt
\includegraphics[width=.33\textwidth]{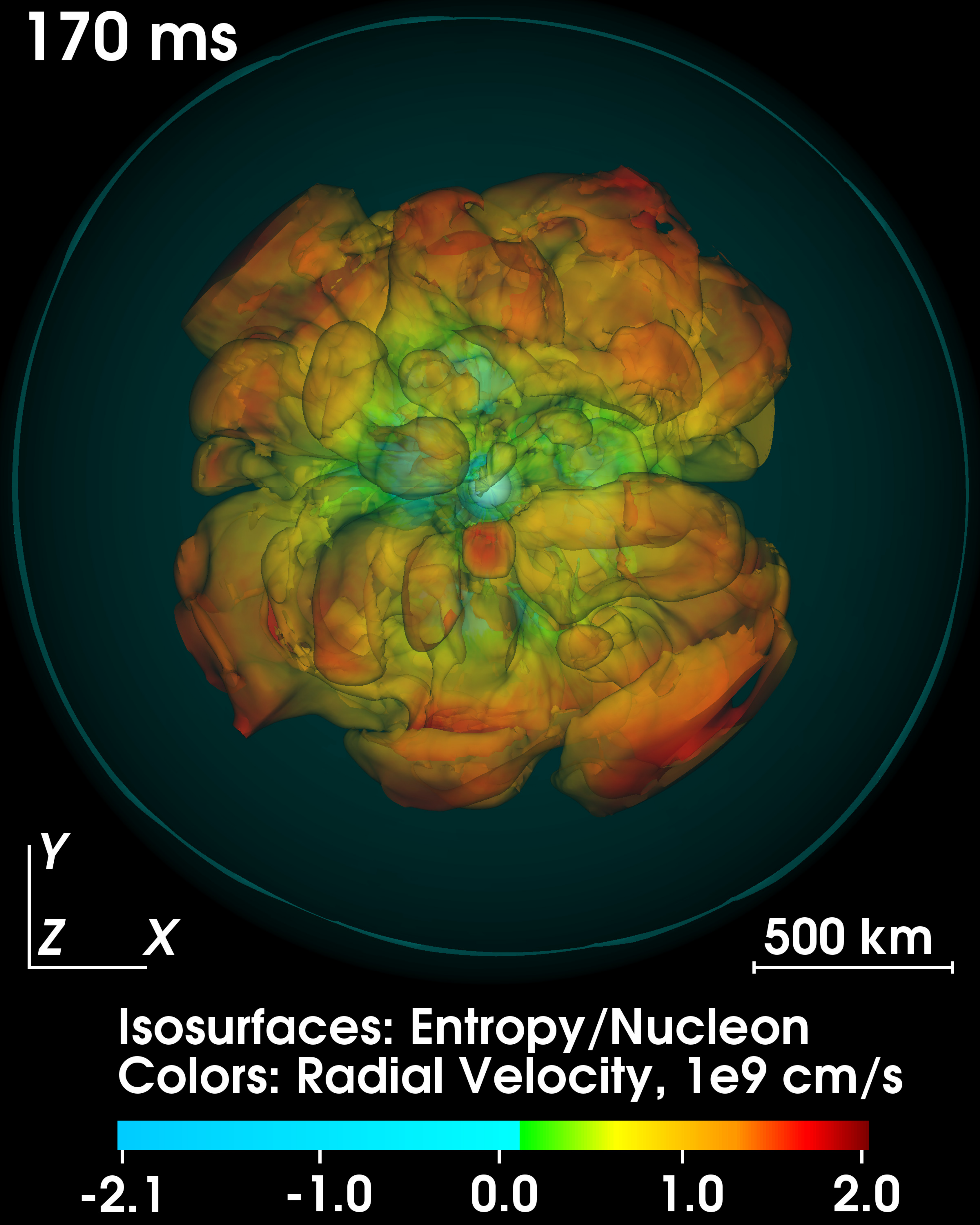}\hskip1.0pt
\includegraphics[width=.33\textwidth]{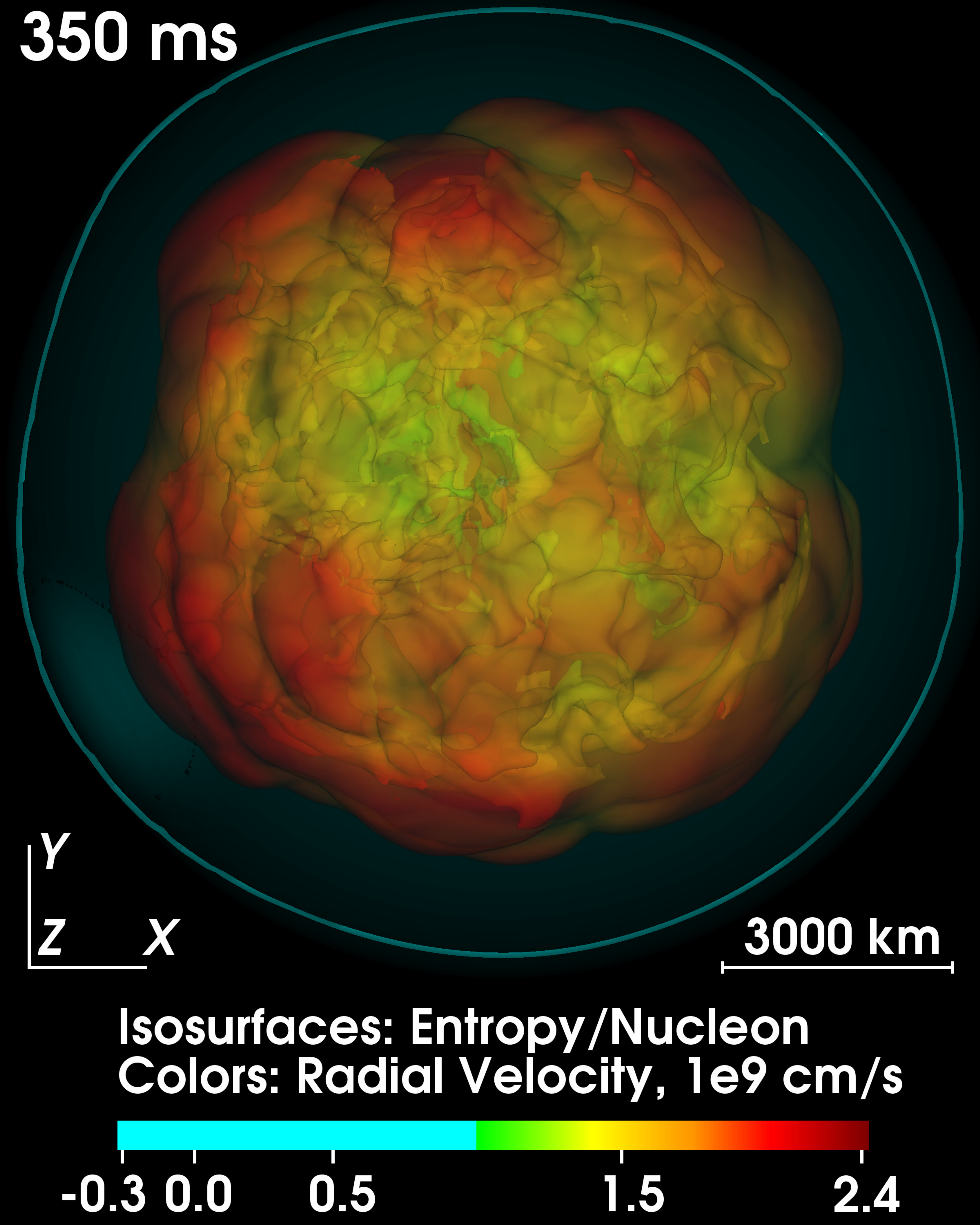}
\caption{
3D iso-entropy surfaces at 90, 170, and 350\,ms after bounce. Colors
represent radial velocities. The supernova shock is visible by a thin
surrounding line, the proto-neutron star by a whitish iso-density surface
of $10^{11}\,$g\,cm$^{-3}$. The yardstick indicates the rapidly growing
volume.
}
\label{fig:3d}
\end{center}
\end{figure*}

\section{Numerical Setup and Progenitor Model}
\label{sec:numerics}

We performed 1D, 2D, and full (4$\pi$)
3D simulations of a nonrotating, zero-metallicity 9.6\,$M_\odot$, iron-core
progenitor provided by A.~Heger (private communication; \citealp{woosley_15}) 
and previously investigated in 2D by \cite{mueller_13}.

We used the \textsc{Prometheus-Vertex} hydrodynamics
code with three-flavor, energy-dependent, ray-by-ray-plus (RbR+) neutrino 
transport including the full set of neutrino reactions and microphysics
\citep{rampp_02,buras_06} applied in 3D
also by \cite{hanke_13} and \cite{tamborra_14},
in particular the high-density equation of state of \cite{lattimer_91}
with a nuclear incompressibility of $K=220$\,MeV. Our simulations were 
conducted with a 1D gravity potential (which is unproblematic for
nearly spherical explosions) including general relativistic 
corrections \citep{marek_06}. For the first time we employed the
newly implemented axis-free Yin-Yang grid \citep{kageyama_04}.
The implementation followed \cite{wongwathanarat_10} and posed no 
particular problems for the RbR+ transport. Conservation laws are 
globally fulfilled with an accuracy of $\sim$10$^{-3}$ over several 
100\,ms for our angular resolution of 2$^\circ$. The radial
grid had a reflecting boundary condition at the coordinate center and
an inflow condition at the outer boundary of $10^9$\,cm. It had 400
nonequidistant zones initially and was refined in steps up to $>$600
zones, providing an increasingly better resolution of 
$\Delta r/r\,\sim$\,0.01\,...\,0.004
around the gain radius. For the neutrino transport 12 geometrically
spaced energy bins with an upper bound of 380\,MeV were used.

\section{Simulation Results}
\label{sec:results}

The relatively steep density gradient above the iron core enables a
neutrino-driven explosion of the 9.6\,$M_\odot$ star at roughly 300\,ms
after bounce even in 1D (left panels in Fig.~\ref{fig:explosion}), similar
to (but more delayed than) oxygen-neon-magnesium-core progenitors exploding 
as ``electron-capture SNe'' \citep{kitaura_06,janka_08,fischer_10}.
The energy of such a late explosion, however, 
remains low, only $E_\mathrm{expl}\sim 2\times 10^{49}$\,erg 
(Fig.~\ref{fig:explosion}),
because it is provided mainly by the recombination of free nucleons
to $\alpha$-particles and iron-group nuclei in the gain layer (see 
\citealp{scheck_06}) and the corresponding mass at $t_\mathrm{pb}>300$\,ms
is only 2.2\,$\times 10^{-3}\,M_\odot$ in 1D. 
Of course, the subsequent neutrino-driven wind 
from the proto-neutron star (PNS) will increase the power of the explosion.
$E_\mathrm{expl}$ denotes the instantaneous ``diagnostic energy'',
which is defined by
\begin{equation}
E_\mathrm{expl}=\int_{e_\mathrm{tot}>0,\mathrm{post shock}}\!\ud V\,\,\rho
e_\mathrm{tot}\,.
\label{eq:expenergy}
\end{equation}
Here, $\rho$ is the density, and the volume integration is performed over
the postshock region where the total specific energy,
\begin{equation}
e_\mathrm{tot}=e+\frac{1}{2}|\textit{\textbf{v}}|^2+\Phi+\left[
e_\mathrm{bind}({}^{56}\mathrm{Fe})-e_\mathrm{bind}\right]\,,
\label{eq:totenergy} 
\end{equation}
is positive, with $e$, $\frac{1}{2}|\textit{\textbf{v}}|^2$, and $\Phi$
being the specific internal, kinetic, and (Newtonian) gravitational energies.
The bracketed term expresses the difference between the nuclear binding 
energies per unit mass of all nucleons finally recombined to iron-group nuclei
and for the nuclear composition at a given time. It therefore accounts for the
maximum release of nuclear binding energy and corresponds to an upper limit
of $E_\mathrm{exp}$, while omitting this term yields a lower bound on
$E_\mathrm{exp}$ (red and blue lines, respectively, in bottom left panel 
of Fig.~\ref{fig:explosion}).
The binding energy of the stellar layers ahead of the shock
plays only a minor role in the energy budget of the explosion because at
$t_\mathrm{pb}=400$\,ms it is only $-3.5\times 10^{48}$\,erg in the 1D
simulation and even only $-9.5\times 10^{47}$\,erg in our multi-dimensional
simulations.

\subsection{Explosion Dynamics and Properties: 2D vs 3D}

\begin{figure}
\includegraphics[width=\columnwidth]{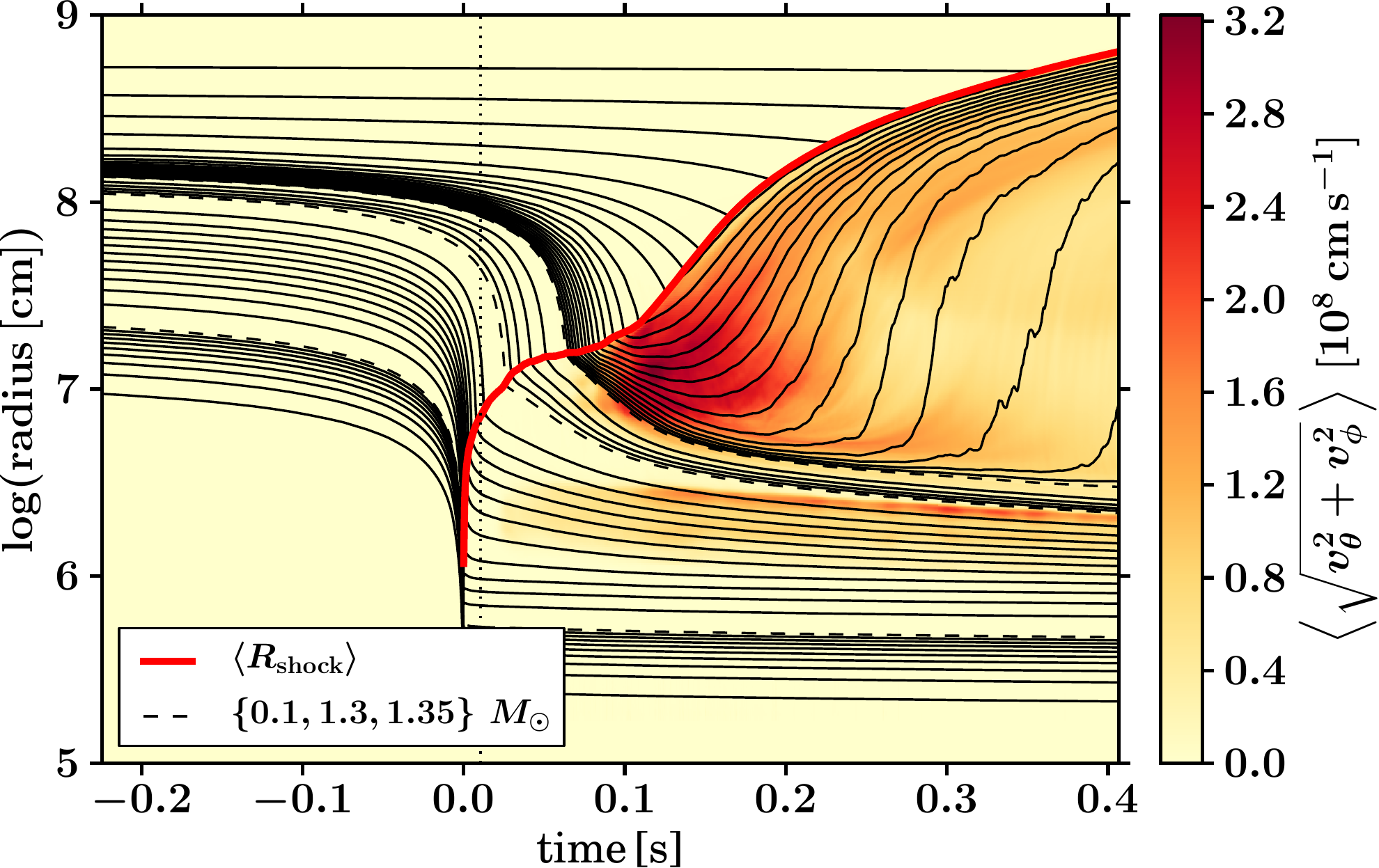}
\caption{Mass-shell (radii for chosen enclosed masses) evolution
of the 3D explosion with colors showing rms values of the
nonradial velocity. The red solid line marks the angle-averaged shock radius,
the dashed lines separate regions with mass spacings of 0.01, 0.1, 0.01, and
0.001\,$M_\odot$ (from inside outwards). The 3D simulation was started from a
1D model at $\sim$10\,ms after bounce (vertical dotted line).
}
\label{fig:mass_shells}
\end{figure}

\begin{figure*}
\plotone{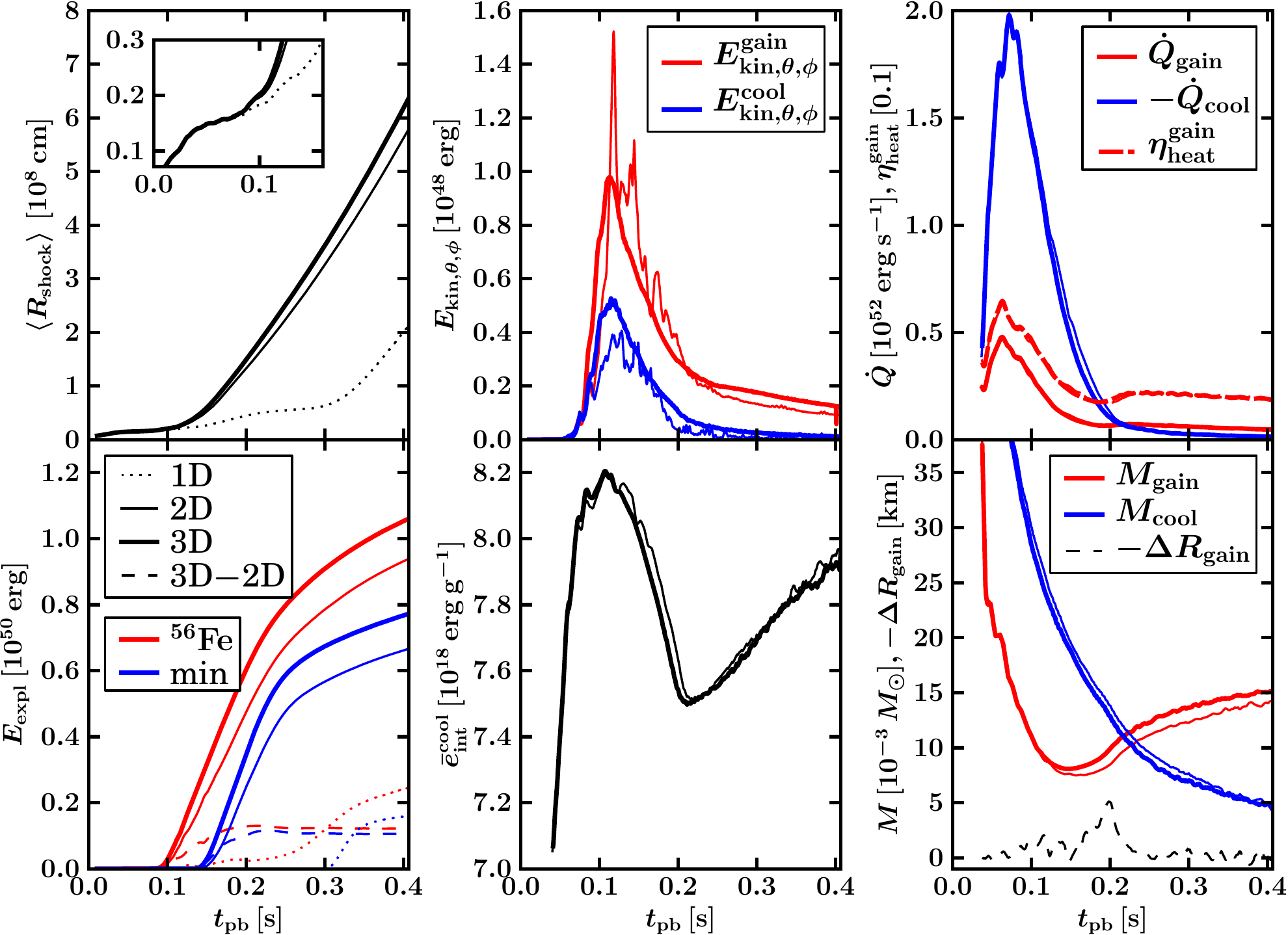}
\caption{Explosion parameters as functions of post-bounce time.
1D results are displayed by dotted lines, 2D by thin solid
lines, and 3D by thick solid lines. Angle-averaged shock radii ({\em upper
left}); explosion energies with upper limits in red, lower limits in blue
(see text) and 3D-2D differences with dashed lines ({\em bottom left}); kinetic
energies of non-radial mass motions in the gain (red) and cooling layers (blue;
{\em upper middle}); average specific internal energies in the cooling layer
({\em lower middle}); total net neutrino-heating rates in the gain (red)
and cooling layers (blue) and neutrino-heating efficiencies (dashed, scaled
by a factor of 10; {\em top right}); masses in the heating (red) and cooling
layers (blue) and difference of angle-averaged gain radii,
$-\Delta R_\mathrm{gain}= -(R_\mathrm{gain}^\mathrm{3D}-R_\mathrm{gain}^\mathrm{2D})$
(dashed; {\em bottom right}).
}
\label{fig:explosion}
\end{figure*}

In 3D convective overturn in the neutrino-heated postshock layer develops
at $t_\mathrm{pb}\gtrsim$\,70\,ms, showing the well-known Rayleigh-Taylor
mushrooms (Fig.~\ref{fig:3d}, left) and increasing nonradial 
velocities (colors in Fig.~\ref{fig:mass_shells}). The postshock convection
reaches its maximum activity between about 100\,ms and 200\,ms after
bounce with a prominent maximum at $\sim$120\,ms, at which time
the shock starts its accelerated expansion (cf.\ Figs.~\ref{fig:mass_shells}
and \ref{fig:explosion}) and the postshock flow becomes highly turbulent
(Fig.~\ref{fig:3d}, middle). At $t_\mathrm{pb}\gtrsim$\,200\,ms the 
shock and postshock matter expand with $\sim$25,000\,km\,s$^{-1}$,
and the ejecta attain a nearly self-similar structure
(Fig.~\ref{fig:3d}, right). This is the time when convection and
turbulence behind the shock become weaker again 
(less intense red in Fig.~\ref{fig:mass_shells}) and mass shells
leaving the PNS surface indicate that the compact remnant begins to lose
material in the low-density, high-entropy baryonic wind driven by 
neutrino heating above the neutrinosphere.

In our multi-dimensional simulations the shock stagnates only for 
$\sim$120--130\,ms and then expands to 
$\left<R_\mathrm{shock}\right>\approx 6000$\,km at 400\,ms postbounce (p.b.)
compared to only $\sim$2000\,km in the 1D case (Fig.~\ref{fig:explosion}).
Angle-averaged quantities $X(r)$ are computed according to
\begin{equation}
\left< X(r)\right>\equiv\frac{\int X(r)\,\ud\Omega}{\int\ud\Omega}\,.
\label{eq:angleave}
\end{equation}
Positive total energies in the postshock layer develop only shortly after
the onset of outward shock acceleration. Interestingly, in 3D the shock
expands faster and the explosion energy increases more steeply
and to a higher value than in 2D. At $t_\mathrm{pb}=400$\,ms the
diagnostic energy reaches 
\begin{equation}
E_\mathrm{expl}^\mathrm{3D}(400\,\mathrm{ms})\sim(0.77\,...\,1.05)\times 
10^{50}\,\mathrm{erg} 
\label{eq:expenergy3d}
\end{equation}
in 3D with an increase at a rate of
$\sim$(0.75\,...\,$1.25)\times 10^{50}$\,erg\,s$^{-1}$
due to the PNS wind for the cases of minimal and maximal energy, respectively,
whereas it is only $\sim$(0.67\,...\,$0.92)\times 10^{50}$\,erg in 2D
with a similar growth rate by the PNS-wind power as in 3D\footnote{Tests
with doubled angular resolution and different seed perturbations revealed 
$\sim$4\% variations of $E_\mathrm{expl}^\mathrm{2D}(400\,\mathrm{ms})$.}.

While the onset of the explosion in multi-dimensions
is aided by buoyant convection,
the steeper rise of the explosion energy in 3D can be understood by
a smaller net energy-loss rate, $\dot Q_\mathrm{cool}$, in the  
neutrino-cooling layer during 100\,ms\,$\lesssim t_\mathrm{pb}\lesssim 200$\,ms
(Fig.~\ref{fig:explosion}). $\dot Q_\mathrm{cool}$ is evaluated between the 
mean gain radius, 
$R_\mathrm{gain}$, and an inner radius $R_0\equiv r(\tau_{\nu_e}=3)$, 
which is (somewhat arbitrarily) defined at an optical depth of
3 for $\nu_e$ (all quantities are evaluated on angle-averaged profiles).
In contrast, the integrated net 
neutrino-heating rate in the gain layer, $\dot Q_\mathrm{gain}$,
and the heating efficiency, $\eta_\mathrm{heat}^\mathrm{gain}=\dot 
Q_\mathrm{gain}(\dot E_{\nu_e}+\dot E_{\bar{\nu}_e})^{-1}$
(with $\dot E_{\nu_i}$ being the direction-averaged luminosities of
$\nu_e$ and $\bar\nu_e$), do not exhibit any
significant 3D-2D differences (Fig.~\ref{fig:explosion}).

The reduced energy-loss rate in the
cooling layer for 100\,ms\,$\lesssim t_\mathrm{pb}\lesssim 200$\,ms
is associated with a systematically lower specific internal energy,
$\bar{e}_{\mathrm{int}}^\mathrm{cool} \equiv 
\left(\int_{R_0}^{R_\mathrm{gain}}\ud V\rho e\right)M_\mathrm{cool}^{-1}$,
averaged over the cooling-layer mass 
$M_{\mathrm{cool}}=\int_{R_0}^{R_\mathrm{gain}}\ud V\rho$
(Fig.~\ref{fig:explosion}). This suggests lower temperatures in the 
cooling layer, less efficient neutrino emission, and therefore an inward
shift of the gain radius in 3D compared to 2D: 
$R_\mathrm{gain}^{3D}<R_\mathrm{gain}^{2D}$ (Fig.~\ref{fig:explosion}).

Consequently, the mass of the gain layer, $M_{\mathrm{gain}} =
\int_{R_\mathrm{gain}}^{\left< R_\mathrm{shock}\right>}\ud V\rho$,
is larger and, correspondingly, the mass in the cooling layer,
$M_{\mathrm{cool}}$, smaller in 3D. While the 3D-2D
difference of $M_{\mathrm{cool}}$ varies with time because the 
definition of $R_0$ depends on model-specific details, the gain-mass
difference grows to 
$\Delta M_\mathrm{gain}\approx 1.2\times 10^{-3}\,M_\odot$ between
100\,ms and 200\,ms and remains essentially constant afterwards.
The nuclear recombination energy of this mass difference,
\begin{equation}
\Delta M_\mathrm{gain}\times e_\mathrm{recomb}\sim(1.6\,...\,2.0)\times
10^{49}\,\mathrm{erg}\gtrsim\Delta E_\mathrm{expl}^{\mathrm{3D}-\mathrm{2D}} 
\label{eq:deexpl}
\end{equation}
for $e_\mathrm{recomb}\sim (7\,...\,8.8)\,\mathrm{MeV/nucleon}
\approx (6.76\,...\,8.49)\times 10^{18}\,$erg\,g$^{-1}$ (depending on whether 
free nucleons recombine to $\alpha$-particles or iron-group elements)
can easily account for the 3D-2D differences of the explosion energy in
Fig.~\ref{fig:explosion}.

Turbulent pressure in the gain layer is understood to be
supportive for the onset of neutrino-driven explosions
\citep{murphy_13,couch_14o,mueller_14b}. However, for our
low-mass SN progenitor this effect cannot be crucial for the 
observed 2D-3D difference. Despite the
faster and more energetic explosion, the 3D model exhibits a {\em lower}
turbulent kinetic energy in the gain layer during the crucial
period 100\,ms\,$\lesssim t_\mathrm{pb}\lesssim\,200$\,ms, visible 
by the nonradial contribution
$E_{\mathrm{kin},\theta,\phi}^\mathrm{gain}=\int_{R_\mathrm{gain}}^{\left< 
R_\mathrm{shock}\right>}\ud V\,\frac{1}{2}\rho (v_\theta^2+v_\phi^2)$
in Fig.~\ref{fig:explosion}. Taking into account the larger mass in the 
gain layer, it means that the nonradial fluid motions are much less 
vigorous in the 3D case. 

In contrast, the nonradial kinetic energy in the cooling layer,
$E_{\mathrm{kin},\theta,\phi}^\mathrm{cool}=\int_{R_0}^{R_\mathrm{gain}} 
\ud V\,\frac{1}{2}\rho (v_\theta^2+v_\phi^2)$, is smaller in the 2D 
simulation despite the higher mass of this layer in 2D. This
suggests that nonradial motions are much more efficiently damped by
dissipation in stronger deceleration shocks and due to symmetry 
constraints (reflective axis boundaries) in the cooling layer of
the 2D model.

\begin{figure*}
\plotone{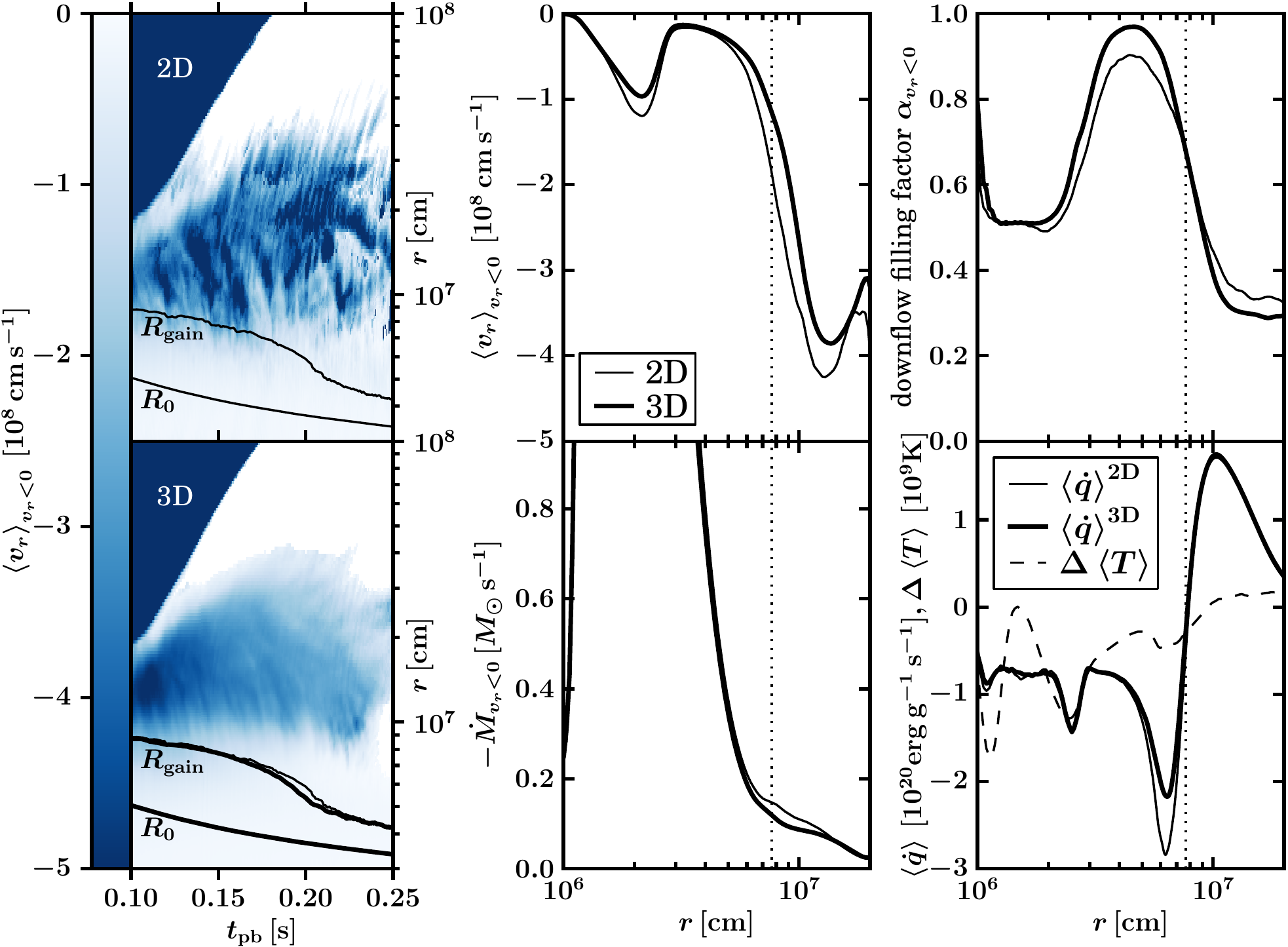}
\caption{Profiles of angle-averaged radial velocities of infalling
($v_r<0$) matter, outgoing shock (sharp blue-white discontinuity), mean gain radius,
$R_\mathrm{gain}$, and inner boundary of cooling layer, $R_0$, as functions
of post-bounce time for 2D ({\em top left}) and 3D models ({\em bottom left}).
Time-averaged (over 0.1\,s$\le t_\mathrm{pb}\le 0.2$\,s) radial profiles of
angle-averaged downflow velocities, $\left< v_r\right>_{v_r<0}$,
(i.e., of matter with $v_r<0$; {\em upper middle});
mass-infall rates in downflows, $\dot{M}_{v_r<0}$ ({\em bottom middle});
``surface filling factors'' of downflows, $\alpha_{v_r<0}$ ({\em upper right});
and angle-averaged specific net neutrino-heating/cooling rates,
$\left<\dot{q}\right>$, and 3D-2D temperature difference,
$\Delta\left<T\right> =\left< T\right>^\mathrm{3D}-\left<T\right>^\mathrm{2D}$
({\em bottom right}). Thin solid lines show 2D, thick solid lines 3D results,
the vertical dotted lines indicate the time-averaged gain radius in 3D.
}
\label{fig:downflows}
\end{figure*}

\begin{figure*}[!]
\includegraphics[width=\textwidth]{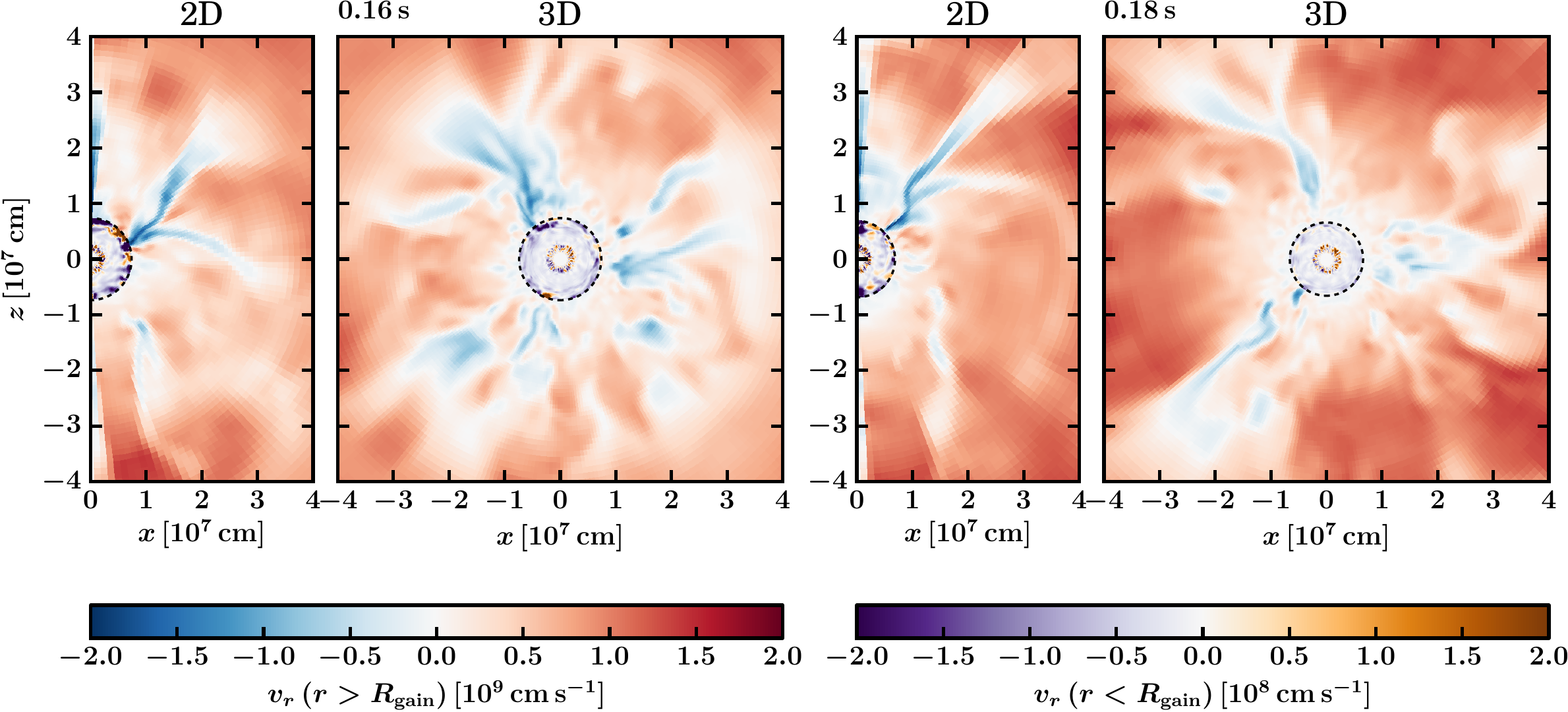}
\caption{Cross-sectional cuts with radial velocities in the
$x$-$z$ plane at 0.16\,s ({\em left}) and 0.18\,s after bounce ({\em right})
for 2D ({\em left semi-panels}) and 3D models ({\rm right semi-panels}).
The mean gain radius, $R_\mathrm{gain}$, is marked by a dashed circle. For
better visibility different color scales are used outside and inside of
$R_\mathrm{gain}$ (left and right color bar, respectively).
}
\label{fig:vex}
\end{figure*}

\subsection{Detailed Analysis of 2D-3D Flow Differences}

The reason for more favorable explosion conditions in 3D becomes clearer 
through radial profiles of the turbulent accretion flow in the
gain and cooling layers (Figs.~\ref{fig:downflows} and \ref{fig:vex}).
Here, angular averages of quantities $X(r)$ over infalling matter, i.e.,
over regions with radial velocity $v_r<0$, are computed according to
\begin{equation}
\left< X(r)\right>_{v_r<0}\equiv
\frac{\int X(r)\,\Theta(-v_r)\,\ud\Omega}{\int\Theta(-v_r)\,\ud\Omega}\,,
\label{eq:angleavevr}
\end{equation}
where $\Theta(x)$ is the heaviside function with $\Theta(x)=1$ for
$x\ge 0$ and $\Theta(x)=0$ otherwise.

Radial velocity profiles of $v_r$, angle-averaged over convective downdrafts
in the postshock accretion layer, exhibit stronger local variations with higher
extrema and pronounced intermittency-like behavior in 2D, whereas the 
angle-averaged infall velocities in the 3D model appear smoother and show
less extreme fluctuations (Fig.~\ref{fig:downflows}, left). This points
to a larger number of smaller-scale convective
structures, thus reducing the statistical variance, and suggests less vigorous
radial mass motions in the 3D model, which is compatible with the lower
nonradial kinetic energy in the gain layer between 100\,ms and 200\,ms
p.b.\ (cf.\ Fig.~\ref{fig:explosion}).

This conclusion is supported by cross-sectional cuts 
(Fig.~\ref{fig:vex}), where narrow convective downflows in 2D 
possess higher velocities, in agreement with radial profiles
of $v_r$ averaged over inflows and time-averaged over
100\,ms\,$\le t_\mathrm{pb}\le 200$\,ms (Fig.~\ref{fig:downflows}).
The higher infall velocities in 2D in the gain layer and a fair
part of the cooling layer are associated with a greater
mass-accretion rate around the gain radius (Fig.~\ref{fig:downflows}).
Moreover, in 2D a larger fraction of the sphere in the gain layer
is subtended by downflows, which is expressed by the ``downflow
filling factor'',
\begin{equation}
\alpha_{v_r<0}\equiv\frac{\int\Theta(-v_r)\,\ud\Omega}{\int\ud\Omega}\,,
\label{eq:surfrac}
\end{equation}
in Fig.~\ref{fig:downflows}. In contrast, in the cooling layer
$\alpha_{v_r<0}^\mathrm{2D} < \alpha_{v_r<0}^\mathrm{3D}$. All
these findings are in line with higher kinetic energies of radial
and nonradial flows in the gain layer and slightly lower nonradial kinetic 
energy in the cooling layer for the 2D model (Fig.~\ref{fig:explosion}).

At first glance, the higher mass-accretion rate around 
$R_\mathrm{gain}$ in 2D seems to contradict the 2D-3D differences of
the flow geometry visible in Fig.~\ref{fig:vex}: In 2D one counts one 
or two prominent convective downdrafts, consistent with a convective
cell pattern characterized by low spherical-harmonics modes, i.e.,
large-scale structures with few downflows. In contrast, many
more downflow funnels exist in 3D, although with smaller infall 
velocities. Nevertheless, the 2D model channels more mass per time
through the gain radius. Obviously, the total volume
of toroidal ``sheets'' formed by axi-symmetric (2D) downflows
is bigger than the volume encompassed by 3D accretion funnels.
Exactly this is expressed by the larger $\alpha_{v_r<0}$ in the gain
layer, which, together with bigger
infall velocities, accounts for the higher mass-accretion rate in 2D.

\subsection{3D Turbulence Facilitating Explosion}
 
The reduced rate and less powerful infall of mass through the gain
radius is therefore the fundamental reason for more favorable explosion
conditions in 3D. Higher downflow velocities in 2D lead to more violent
deceleration of the accretion downdrafts in shocks
as they penetrate into the cooling layer (see Fig.~\ref{fig:vex} and
the more dramatic decrease of $|\left< v_r\right>_{v_r<0}|$ 
in 2D in Fig.~\ref{fig:downflows}). On the one hand this causes stronger
gravity-wave activity below $R_\mathrm{gain}$, which can be recognized by
larger angular variations of $v_r$ for the 2D model in this region
in Fig.~\ref{fig:vex}. On the other hand the impact of downflows dissipates
kinetic energy, for which reason the 2D simulation shows a lower kinetic
energy of nonradial mass motions in the cooling layer despite its higher
nonradial kinetic energy in the heating layer (Fig.~\ref{fig:explosion}).
More important, however, is the enhanced shock heating of the
cooling layer, which leads to higher temperatures in this region in 2D
(Fig.~\ref{fig:downflows}). Estimates confirm that this difference of the
angle-averaged temperature of several $10^9$\,K (or several percent of the
local angle-averaged temperature) is responsible for the larger net cooling 
rate of the 2D model just below $R_\mathrm{gain}$ (Fig.~\ref{fig:downflows}), 
because local temperature
fluctuations are considerably bigger than the relative differences of 
the angular averages and neutrino emission rates by $e^\pm$-captures
and $e^+e^-$-pair annihilation scale with $T^6$ and $T^9$, respectively.
Consequently, one finds 
$R_\mathrm{gain}^\mathrm{2D}>R_\mathrm{gain}^\mathrm{3D}$ 
during 170\,ms\,$\lesssim t_\mathrm{pb}\lesssim$\,220\,ms
(Fig.~\ref{fig:downflows}, bottom left).
Since the temperature in the gain layer is only marginally higher in 3D,
the net heating rates between $R_\mathrm{gain}$ and shock are hardly
distinguishable for 2D and 3D (Fig.~\ref{fig:downflows}, bottom right).

The 2D-3D differences in the convective postshock layer, 
in particular higher downflow velocities and larger mass-infall
rate in 2D, can be attributed to well-known differences
of 2D and 3D turbulence discussed previously (e.g.,
\citealp{hanke_12,burrows_12,dolence_13,couch_13,couch_14,dolence_13,abdikamalov_14}):
Turbulent energy cascading leads to fragmentation of 
large-scale flows to smaller-scale vortices in 3D, whereas the 
inverse energy cascade in 2D tends to enhance kinetic energy
on the largest possible scales. The turbulent fragmentation (e.g.\
through Kelvin-Helmholtz instability) of 
the postshock accretion flow affects the flow characteristics
most strongly near the gain radius, for which
reason 2D-3D differences of $\dot{M}_{v_r<0}$ increase as the 
infalling gas approaches $R_\mathrm{gain}$ (Fig.~\ref{fig:downflows}).
Because of enhanced turbulent energy in small-scale vortices,
the 3D flow becomes less coherent, and turbulent small-scale
structures keep mass in the gain layer and reduce $\dot{M}_{v_r<0}$,
while in 2D the mass-infall rate continues to grow all the way 
towards $R_\mathrm{gain}$.

\section{Conclusions}
\label{sec:conclusions}

Performing simulations of neutrino-driven explosions for a
9.6\,$M_\odot$ star with the \textsc{Prometheus-Vertex} code
in 1D and multi-D, we found significant 2D-3D differences of the 
accretion flow in the convective postshock layer. Turbulent 
fragmentation to small-scale vortices keeps more matter in the
gain layer in 3D and decreases the rate and velocity of mass 
accretion into the cooling layer. This reduces the dissipation 
of kinetic energy in the cooling region, lowers the temperature
there and decreases the energy loss by neutrino emission. The 
corresponding inward shift of the gain radius leads to an 
increase of the mass in the gain layer, whose higher nucleon
recombination energy accelerates shock expansion in 3D and 
boosts the explosion energy by $\gtrsim$10\% to $>$10$^{50}$\,erg
with a steeper rise at the beginning of the explosion.

We thus presented the first successful neutrino-driven explosion
in a fully self-consistent 3D simulation and, for the first time, 
identified effects that foster and strengthen
neutrino-driven explosions in 3D compared to 2D.
In contrast to the unfavorable influence of 3D turbulence
observed in previous studies of more massive stars (e.g.,
\citealp{hanke_12,couch_13,couch_14,takiwaki_12,takiwaki_14,abdikamalov_14}),
the decrease
of the mass inflow rate through the gain radius and reduced
neutrino-energy loss in the cooling layer below $R_\mathrm{gain}$
have a healthy influence on the explosion in our 9.6\,$M_\odot$ 
progenitor in 3D.

It is difficult to speculate whether these effects could play an
important role also for more massive stars, in particular when
convection dominates the postshock dynamics \citep{burrows_12,murphy_13}
and current numerical resolution deficiencies (see, e.g., discussions
by \citealp{radice_15,abdikamalov_14,hanke_12}) have been 
overcome. We emphasize that our present grid resolution is 
insufficient to achieve convergence and to capture fully developed
turbulence. However, we attribute our 2D-3D differences to generic
differences of convective downflow dynamics, which are likely to
remain valid even for higher resolution. We have identified turbulent
fragmentation of 3D flows in the gain layer as helpful for stronger 
explosions than in 2D, in sharp contrast to disadvantageous
consequences of 3D effects for the onset of explosions reported
for more massive progenitors in previous works. Higher resolution
might even enhance our described effects and their
explosion-facilitating consequences rather than endangering them.

\acknowledgements
We thank R.~Bollig, J.~von Groote, F.~Hanke, B.~M{\"u}ller, 
E.~M{\"u}ller, A.~Wongwathanarat, E.~Erastova, and
M.~Rampp for support and discussions and J.~Guilet for careful
reading and comments.
Funding by Deutsche Forschungsgemeinschaft through grants SFB/TR7
and EXC 153 and by the European Union through grant ERC-AdG
No.~341157-COCO2CASA and computing time from the European
PRACE Initiative on SuperMUC (GCS@LRZ, Germany) and CURIE 
(GENCI@CEA, France) are acknowledged. Postprocessing was done 
on Hydra of Rechenzentrum Garching.

\bibliographystyle{apj}

\begin{thebibliography}{}

\bibitem[{{Abdikamalov} {et~al.}(2014){Abdikamalov}, {Ott}, {Radice}, {Roberts},
    {Haas}, {Reisswig}, {Moesta}, {Klion} \& {Schnetter}}]{abdikamalov_14}
    {Abdikamalov}, E., {Ott}, C.D., {Radice}, D., {Roberts}, L.F., {Haas}, R.,
    {Reisswig}, C., {Moesta}, P., {Klion}, H., \& {Schnetter}, E.\ 2014, 
    eprint arXiv:1409.7078

\bibitem[{{Blondin} {et~al.}(2003){Blondin}, {Mezzacappa}, \& {DeMarino}}]{blondin_03}
    {Blondin}, J.M., {Mezzacappa}, A., \& {DeMarino}, C.\ 2003, \apj, 584, 971

\bibitem[{{Bruenn} {et~al.}(2013){Bruenn}, {Mezzacappa}, {Hix}, {Lentz},
    {Bronson}, {Lingerfelt}, {Blondin}, {Endeve}, {Marronetti} \& {Yakunin}}]{bruenn_13}
    {Bruenn}, S.W., {Mezzacappa}, A., {Hix}, W.R, {Lentz}, E.J., {Bronson}, M.O.E.,
    {Lingerfelt}, E.J., {Blondin}, J.M., {Endeve}, E., {Marronetti}, P., \&
    {Yakunin}, K.N.\ 2013, \apjl, 767, L6 

\bibitem[{{Bruenn} {et~al.}(2014){Bruenn}, {Lentz}, {Hix}, {Mezzacappa}, {Harris},
    {Bronson}, {Endeve}, {Blondin}, {Chertkow}, {Lingerfelt}, {Marronetti} \& 
    {Yakunin}}]{bruenn_14}
    {Bruenn}, S.W., {Lentz}, E.J., {Hix}, W.R, {Mezzacappa}, A., {Harris}, J.A.,
    {Bronson}, M.O.E., {Endeve}, E., {Blondin}, J.M., {Chertkow}, M.A., 
    {Lingerfelt}, E.J., {Marronetti}, P., \& {Yakunin}, K.N.\ 2014, eprint arXiv:1409.5779 

\bibitem[{{Buras} {et~al.}(2006){Buras}, {Rampp}, {Janka}, \& {Kifonidis}}]{buras_06}
    {Buras}, R., {Rampp}, M., {Janka}, H.-T., \& {Kifonidis}, K.\ 2006, \aap, 447, 1049

\bibitem[{{Burrows} {et~al.}(1995){Burrows}, {Hayes}, \& {Fryxell}}]{burrows_95}
    {Burrows}, A., {Hayes}, J., \& {Fryxell}, B.A.\ 1995, \apj, 450, 830

\bibitem[{{Burrows} {et~al.}(2012){Burrows}, {Dolence}, \& {Murphy}}]{burrows_12}
    {Burrows}, A., {Dolence}, J.C., \& {Murphy}, J.W.\ 2012, \apj, 759, 5

\bibitem[{{Couch}(2013)}]{couch_13} {Couch}, S.M.\ 2013, \apj, 775, 35

\bibitem[{{Couch} \& {O'Connor}(2014)}]{couch_14} {Couch}, S.M. \& {O'Connor}, E.P.\
    2014, \apj, 785, 123

\bibitem[{{Couch} \& {Ott}(2013)}]{couch_13o} {Couch}, S.M. \& {Ott}, C.D.\
    2013, \apjl, 778, L7  

\bibitem[{{Couch} \& {Ott}(2014)}]{couch_14o} {Couch}, S.M. \& {Ott}, C.D.\
    2014, eprint arXiv:1408.1399 

\bibitem[{{Dolence} {et~al.}(2013){Dolence}, {Burrows}, {Murphy}, \&
    {Nordhaus}}]{dolence_13} {Dolence}, J.C., {Burrows}, A., {Murphy}, J.W., \&
    {Nordhaus}, J.\ 2013, \apj, 765, 110

\bibitem[{{Fischer} {et~al.}(2010){Fischer}, {Whitehouse}, {Mezzacappa},
    {Thielemann}, \& {Liebend{\"o}rfer}}]{fischer_10}
    {Fischer}, T., {Whitehouse}, S.C., {Mezzacappa}, A., {Thielemann}, F.-K., 
    \& {Liebend{\"o}rfer}, M.\ 2010, \aap, 517, A80

\bibitem[{{Hanke} {et~al.}(2012){Hanke}, {Marek}, {M{\"u}ller}, \& {Janka}}]{hanke_12}
    {Hanke}, F., {Marek}, A., {M{\"u}ller}, B., \& {Janka}, H.-T.\ 2012, \apj, 755, 138

\bibitem[{{Hanke} {et~al.}(2013){Hanke}, {M{\"u}ller}, {Wongwathanarat},
    {Marek} \& {Janka}}]{hanke_13}
    {Hanke}, F., {M{\"u}ller}, B., {Wongwathanarat}, A., {Marek}, A.,
    \& {Janka}, H.-T.\ 2013, \apj, 770, 66 

\bibitem[{{Herant} {et~al.}(1994){Herant}, {Benz}, {Hix}, {Fryer} \& {Colgate}}]{herant_94}
    {Herant}, M., {Benz}, W., {Hix}, W.R., {Fryer}, C.L., \& {Colgate}, S.A.\ 1994, \apj, 
    435, 339

\bibitem[{{Janka} \& {M{\"u}ller}(1996)}]{janka_96} {Janka}, H.-Th. \& {M{\"u}ller}, E.\ 
    1996, \aap, 306, 167  

\bibitem[{{Janka} {et~al.}(2008){Janka}, {M{\"u}ller}, {Kitaura} \& {Buras}}]{janka_08}
    {Janka}, H.-T., {M{\"u}ller}, B., {Kitaura}, F.S., \& {Buras}, R.\ 2008, \aap, 485, 199

\bibitem[{{Kageyama} \& {Sato}(2004)}]{kageyama_04} {Kageyama}, A., \& {Sato}, T.\ 2004,
    Geochem.~Geophys.~Geosys., 5

\bibitem[{{Kitaura} {et~al.}(2006){Kitaura}, {Janka}, \& {Hillebrandt}}]{kitaura_06}
    {Kitaura}, F.S., {Janka}, H.-T., \& {Hillebrandt}, W.\ 2006, \aap, 450, 345

\bibitem[{{Lattimer} \& {Swesty}(1991)}]{lattimer_91} {Lattimer}, J.~M., \&
    {Swesty}, F.~D.\ 1991, Nucl.~Phys.~A, 535, 331

\bibitem[{{Marek} \& {Janka}(2009)}]{marek_09} {Marek}, A. \& {Janka}, H.-T.\ 2009,
    \apj, 694, 664

\bibitem[{{Marek} {et~al.}(2006){Marek}, {Dimmelmeier}, {Janka},
    {M{\"u}ller}, \& {Buras}}]{marek_06}
    {Marek}, A., {Dimmelmeier}, H., {Janka}, H.-T., {M{\"u}ller}, E., \& {Buras},
    R.\ 2006, \aap, 445, 273

\bibitem[{{Mezzacappa} {et~al.}(2015){Mezzacappa}, {Bruenn}, {Lentz}, {Hix}, {Harris},
    {Bronson}, {Endeve}, {Blondin}, {Marronetti} \& {Yakunin}}]{mezzacappa_15}
    {Mezzacappa}, A., {Bruenn}, S.W., {Lentz}, E.J., {Hix}, W.R, {Harris}, J.A.,
    {Bronson}, M.O.E., {Endeve}, E., {Blondin}, J.M., 
    {Marronetti}, P., \& {Yakunin}, K.N.\ 2015, eprint arXiv:1501.01688

\bibitem[{{M{\"u}ller} \& {Janka}(2014a)}]{mueller_14a} {M{\"u}ller}, B. \& {Janka}, H.-T.\ 
    2014a, \apj, 788, 82

\bibitem[{{M{\"u}ller} \& {Janka}(2014b)}]{mueller_14b} {M{\"u}ller}, B. \& {Janka}, H.-T.\
    2014b, eprint arXiv:1409.4783 

\bibitem[{{M{\"u}ller} {et~al.}(2012a){M{\"u}ller}, {Janka}, \&
    {Heger}}]{mueller_12a}
    {M{\"u}ller}, B., {Janka}, H.-T., \& {Heger}, A.\ 2012a, \apj, 761, 72

\bibitem[{{M{\"u}ller} {et~al.}(2012b){M{\"u}ller}, {Janka}, \&
    {Marek}}]{mueller_12b}
    {M{\"u}ller}, B., {Janka}, H.-T., \& {Marek}, A.\ 2012b, \apj, 756, 84

\bibitem[{{M{\"u}ller} {et~al.}(2013){M{\"u}ller}, {Janka}, \&
    {Marek}}]{mueller_13}
    {M{\"u}ller}, B., {Janka}, H.-T., \& {Marek}, A.\ 2013, \apj, 766, 43

\bibitem[{{Murphy} \& {Burrows}(2008)}]{murphy_08} {Murphy}, J.W. \& {Burrows}, A.\ 2008,
    \apj, 688, 1159

\bibitem[{{Murphy} {et~al.}(2013){Murphy}, {Dolence}, \&
    {Burrows}}]{murphy_13} {Murphy}, J.W., {Dolence}, J.C., \& {Burrows}, A.\
    2013, \apj, 771, 52

\bibitem[{{Nordhaus} {et~al.}(2010){Nordhaus}, {Burrows}, {Almgren}, \& 
    {Bell}}]{nordhaus_10} {Nordhaus}, J., {Burrows}, A., {Almgren}, A., \&
    {Bell}, J.\ 2010, \apj, 720, 694 

\bibitem[{{Radice} {et~al.}(2015){Radice}, {Couch}, \&
    {Ott}}]{radice_15}
    {Radice}, D., {Couch}, S.M., \& {Ott}, C.D.\ 2015, eprint arXiv:1501.03169

\bibitem[{{Rampp} \& {Janka}(2002)}]{rampp_02} {Rampp}, M. \& {Janka}, H.-T.\ 2002,
    \aap, 396, 361

\bibitem[{{Scheck} {et~al.}(2006){Scheck}, {Kifonidis}, {Janka}, \& {M{\"u}ller}}]{scheck_06}
    {Scheck}, L., {Kifonidis}, K., {Janka}, H.-T., \& {M{\"u}ller}, E.\ 2006, \aap, 457, 963

\bibitem[{{Suwa} {et~al.}(2013){Suwa}, {Takiwaki}, {Kotake}, {Fischer}, {Liebend{\"o}rfer},
    \& {Sato}}]{suwa_13}
    {Suwa}, Y., {Takiwaki}, T., {Kotake}, K., {Fischer}, T., 
    {Liebend{\"o}rfer}, M., \& {Sato}, K.\ 2013, \apj, 764, 99

\bibitem[{{Takiwaki} {et~al.}(2012){Takiwaki}, {Kotake}, \&
    {Suwa}}]{takiwaki_12}
    {Takiwaki}, T., {Kotake}, K., \& {Suwa}, Y.\ 2012, \apj, 749, 98

\bibitem[{{Takiwaki} {et~al.}(2014){Takiwaki}, {Kotake}, \&
    {Suwa}}]{takiwaki_14}
    {Takiwaki}, T., {Kotake}, K., \& {Suwa}, Y.\ 2014, \apj, 786, 83

\bibitem[{{Tamborra} {et~al.}(2014){Tamborra}, {Hanke}, {Janka}, {M{\"u}ller},
    {Raffelt} \& {Marek}}]{tamborra_14}
    {Tamborra}, I., {Hanke}, F., {Janka}, H.-T., {M{\"u}ller}, B.,
    {Raffelt}, G.G., \& {Marek}, A.\ 2014, \apj, 792, 96

\bibitem[{{Wongwathanarat} {et~al.}(2010){Wongwathanarat}, {Hammer}, \&
    {M{\"u}ller}}]{wongwathanarat_10}
    {Wongwathanarat}, A., {Hammer}, N.J., \& {M{\"u}ller}, E.\ 2010, \aap, 514, 48

\bibitem[{{Woosley} \& {Heger}(2015)}]{woosley_15} {Woosley}, S.E. \& {Heger}, A.\
    2015, \apj, submitted

\end{thebibliography}

\end{document}